\documentclass[journal,10pt]{IEEEtran}
\usepackage{cite}
\usepackage[font=small]{caption}
\usepackage{ifpdf}
\usepackage{epstopdf}
\usepackage{authblk}
\usepackage{commath}
\usepackage{cuted} 
\usepackage{stix}
\usepackage{pgfplots}
\usetikzlibrary{arrows,decorations.markings}
\usepackage{bigints}
\newif\ifCLASSOPTIONonecolumn       \CLASSOPTIONonecolumnfalse
\newif\ifCLASSOPTIONtwocolumn       \CLASSOPTIONtwocolumntrue
\ifCLASSINFOpdf
\usepackage{multicol}
\usepackage{subcaption}
\usepackage{amsmath}
\usepackage{dsfont}
\usepackage{bbold}
\usepackage{amssymb}
\usepackage{algorithmic}
\usepackage{array}
\usepackage{stfloats}
\fnbelowfloat
\usepackage{bigints}
\begin{document}
%

\title{On the Usage of Networked Tethered Flying Platforms for Massive Events -\\ Case Study: Hajj Pilgrimage}
%
%
%

\author[ ]{Baha Eddine Youcef~Belmekki}
\author[ ]{Mohamed-Slim~Alouini}
\affil[ ]{Computer, Electrical and Mathematical
Sciences and Engineering (CEMSE) Division, King Abdullah University of Science and Technology (KAUST), Kingdom of Saudi Arabia,}
\affil[ ]{email: \{bahaeddine.belmekki, slim.alouini\}@kaust.edu.sa}
\setcounter{Maxaffil}{0}
\renewcommand\Affilfont{\small}

\markboth{}%
{Shell \MakeLowercase{\textit{et al.}}: Bare Demo of IEEEtran.cls for IEEE Journals}

{}
\maketitle

\IEEEpeerreviewmaketitle

\begin{abstract}
The Hajj is a religious Muslim pilgrimage undertaken annually by 2—3 million people in Makkah. Consequently, several problems arise due to the sheer number of pilgrims, and therefore negatively impact their stay and the conduct of the rituals. During the Hajj, several problems occur related to mobility, security, and connectivity. The current solutions used to deal with these problems have limitations and they usually require a lot of resources with suboptimal results. In this paper, we proposed an aerial-based solution that rely on Networked Tethered Flying Platforms (NTFPs). NTFPs are flying vehicles such as drones, Helikites, and blimps, that are tethered to the ground via a cable that supplies them with constant data and power. NTFPs can fly at high altitude with a great backhaul capacity and large coverage. We show in this paper how NTFP-based solution solve mobility, security, and connectivity problems during the Hajj and the main advantages and benefits as well as the cost-efficiency of such solution. For the sake of completeness, we also present other similar case studies in which NTFPs can be used.

\end{abstract}
%

\IEEEpeerreviewmaketitle

\section{Introduction}
\subsection{Hajj Pilgrimage}
Hajj is an annual religious pilgrimage to Makkah in Saudi Arabia undertaken each year by 2--3 million people. The specific schedules, dates, and times of the Hajj are set in a specific month in the islamic lunar calendar. The Umrah, on the other hand, is an Islamic pilgrimage to Makkah that can be undertaken at any time. Hence, a large number of pilgrims gather in Saudi Arabia each year to undertake these Islamic rituals. Consequently, due to the massive number of pilgrims, several problems arise, and can negatively impact the conduct of these rituals as well as their stay during the pilgrimage \cite{britannica}.

\begin{figure}[ht!]
\centering
\includegraphics[scale=0.12]{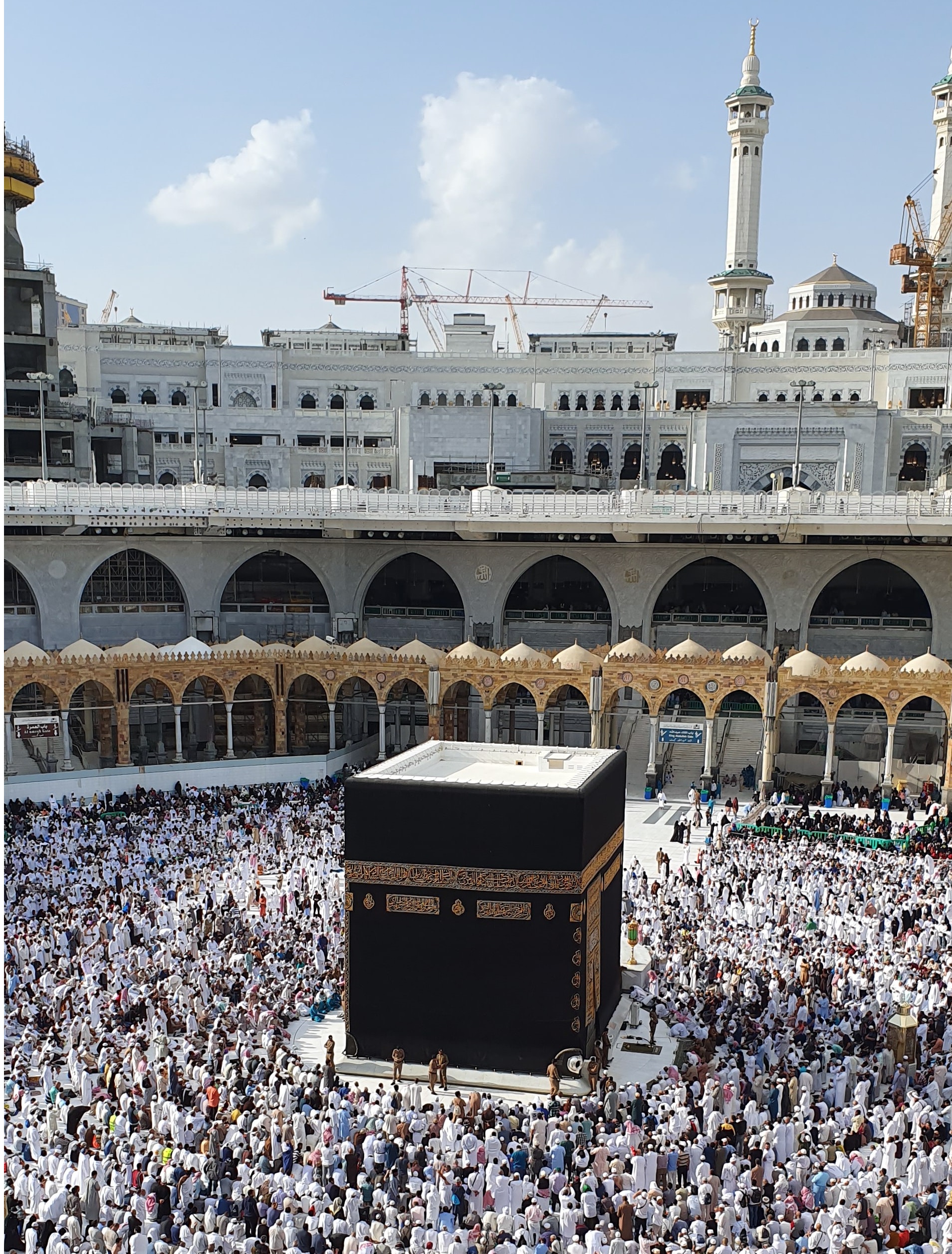} 
\caption{Pilgrims walking around Kaaba in Makkah during the Hajj pilgrimage.}
 \label{}
\end{figure}


\subsection{Problems During the Hajj}
During the Hajj several problems arise related to mobility, security, and connectivity. In the following, we will we detail each of these problems and their impact on the Hajj.
\subsubsection{Mobility Problems}

Several problems can arise from mobility. It can be the poor management that causes a delay, it can be a traffic congestion that can accumulate and hinder the mobility of the vehicles, or it can also be the lack and poor accident management of the traffic. In the context of Hajj, this can be more amplified because of the large number of pilgrims that use transportation in a short period of time.
Also, poor mobility management is often an overlooked problem. The lack or poor mobility management provoke an incremental delay that creates traffic congestion. Consequently, it will cause financial loses to the government due to inadequate management of public transportation. Furthermore, it will impact negatively the experience of the pilgrims using these transportation during the Hajj.
On the other hand, traffic congestion is a serious issue, as the Saudi economist Ihsan Buhaliga said \textit{"If we estimate the cost of every minute delayed on the roads is one riyal, we will find that we are wasting tens of millions of riyals a day and billions a year on the road”} \cite{saudigazette}. Finally, accident avoidance and prevention techniques are saving millions lives each year. Hence it is extremely important to develop further this technique not only to avoid accidents, but to efficiently handle accidents when they happen not only during the Hajj, but all year long.

\begin{figure*}[ht!]
\centering
\includegraphics[scale=0.32]{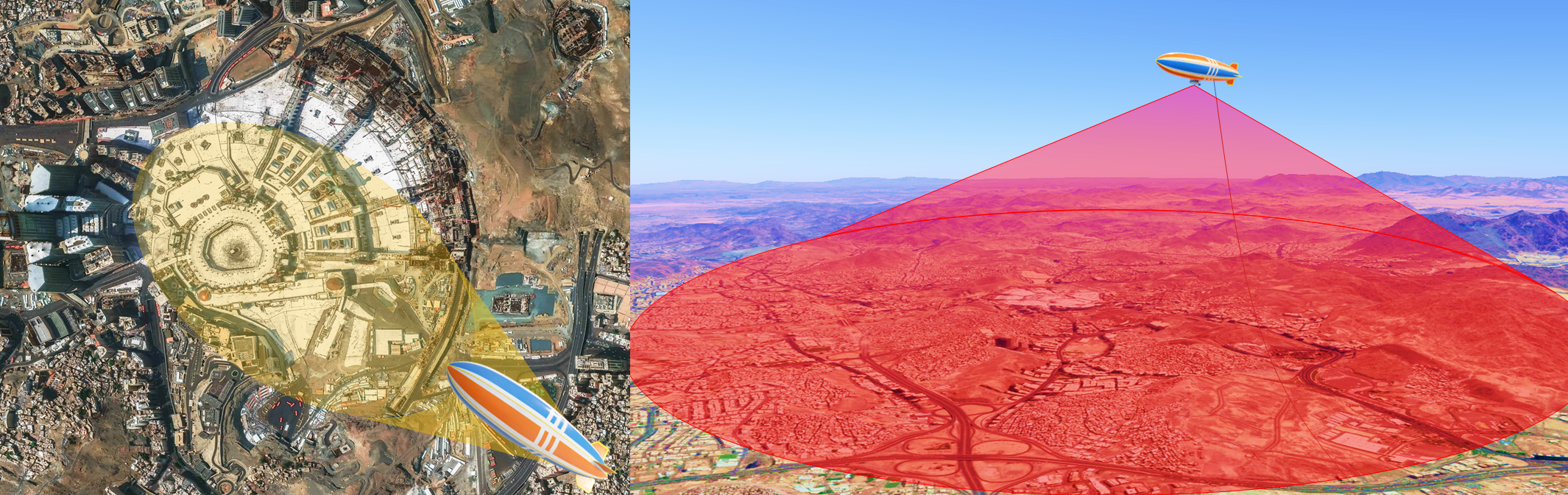} 
\caption{The coverage provided by a tethered blimp over Makkah.}
 \label{}
\end{figure*}

\subsubsection{Security Problems}

Security is vital issue, whether it is homeland security or population security. This is more relevant in Makkah since a large number of pilgrims come each year to undertake religious rituals. Security can involve the surveillance of the border to prevent illegal immigrant or illegal pilgrims. In fact, illegal immigrants have always been a serious problem for several countries that cost millions of dollars each year, and it is very hard to apply a surveillance for borders when they are very long. In addition of the illegal immigrants, there are also illegal pilgrims who try to go on the Hajj without a Hajj certificate. Hence, the importance to apply border control on illegal immigrant and pilgrims.
Security can also involve the surveillance of the pilgrims during the religious rituals, but also outside their religious rituals. Moreover, all the pilgrims are jammed in a single and relatively narrow location which can result in a stampede. The most disastrous stampede happened on 24 September 2015 in Mina causing the deaths of more than 2000 pilgrims.
Two large groups of pilgrims heading to the Jamaraat Bridge intersected from different directions onto the same street. Consequently, a lot of accident and problems can happen which highlight the important of surveillance, especially around the Kaaba. Indeed, a group of armed extremist insurgents attacked Makkah on 20 November 1979 known as the grand mosque seizure. This shows the importance of surveillance during the Hajj.

\subsubsection{Connectivity Problems}
Connectivity is ubiquitous in our everyday lives. However, connectivity can face a lot of problems, such as, the loss of coverage during firefighters' interventions, the loss of coverage during disasters, low data rates, and lack of seamless connectivity during peak hours. During Hajj, pilgrims should be able to benefits from a an excellent quality of internet connection for different purposes such as video calls and mobile Hajj-related applications. 
Indeed, several applications are available to pilgrims to help them perform the rituals \cite{khan2018analytical}. However, according to \cite{majrashi2018user}, many reported problems with Hajj-related apps due to mobile coverage. As an example,one pilgrim and his mother and his sister (70 years old and 45 years old respectively) used a map-app to return to their tents. But, the map-app kept repeatedly changing and updating paths due to outage in connectivity. Consequently, they spent nearly four hours searching for their tent. Because of the long walk, the mother and the sister were unstable for three days since they both had diabetes. This shows how connectivity can impact the quality of the pilgrimage, and in some instances, negatively impact the Pilgrims' health.
To alleviate the lack and poor connection, tower masts have to be erected in the area.  However, tower mats are expensive and erect them take a lot of time. Plus, sometimes the terrain may not be suitable for erecting tower masts. In addition, in the case of a natural disaster, tower masts are often destroyed or non-functional. This can be more harmful for firefighters during their interventions when they must coordinate themselves during a risky and dangerous operation.
Hence it is important to have backups solution in these scenarios, especially to help the first responder and the rescue teams to communicate during the aftermath \cite{alyami2020disaster}.

\subsection{Current Solutions}

The current solutions to the aforementioned problems have some limitations which can require a lot of resources without obtain optimal results.
The current solutions regarding mobility problems rely solely on cameras in building or on poles. However, cameras have limited coverage and mounted at a relatively low altitude. Thus, increasing the number of cameras to cover a given area which makes it difficult to coordinate all these cameras. Also, the monitoring is human-based, hence the surveillance is prone to errors and human bias resulting in missing and overlooking events, hence, not taking the relevant action at the right moment.
Security, on the other hand, depends on cameras and on-site agents. The limitation of cameras has been discussed. The limitations of using agents are also related to human perception and lack of reach. For instance, in border surveillance the agents cannot be deployed along all the border. 
Solutions for connectivity issues are either built several tower masts which are extremely expensive, or rely on flying drones if communication infrastructures are damaged or destroyed by a disaster. However, drones can only fly for limited period of time, ranging for 20 minutes to 40 minutes \cite{thetealmango}.

\subsection{NTFP Solutions}
To tackle the aforementioned problems and overcome the current limitations, we propose a solution in the Hajj context that relies on Networked Tethered Flying Platforms (NTFPs).
NTFPs are flying vehicles that are tethered to the ground via a tether that supply them with constant data and power \cite{belmekki2020unleashing}.
NTFPs can fly at high altitude which allow them to monitor vehicles with a superior coverage compared to camera on the ground. The large coverage offered by the camera mounted on the platform can help monitor the traffic allowing accident prevention, regulating the traffic, and detecting illegal parking.
NTFPs are extremely efficient for surveillance operations. Flying at high altitude allow them to watch the borders with a coverage of several kilometers. They can also detect illegal immigrants from a very long distance, and thus, helping the border control agent to take adequate actions. Also, their high altitude gives them a better and clear view on crowd gatherings. Thus, facilitating the detection of incidents and suspect behaviors.
Finally, NTFPs offer a huge coverage of connectivity and seamless connection thanks to the altitude and their continuous supply in data and power. They can also be deployed very quickly in case of disasters and provide a connection to the rescue team and firefighter teams.

\begin{figure}[t!]
\centering
\includegraphics[scale=0.31]{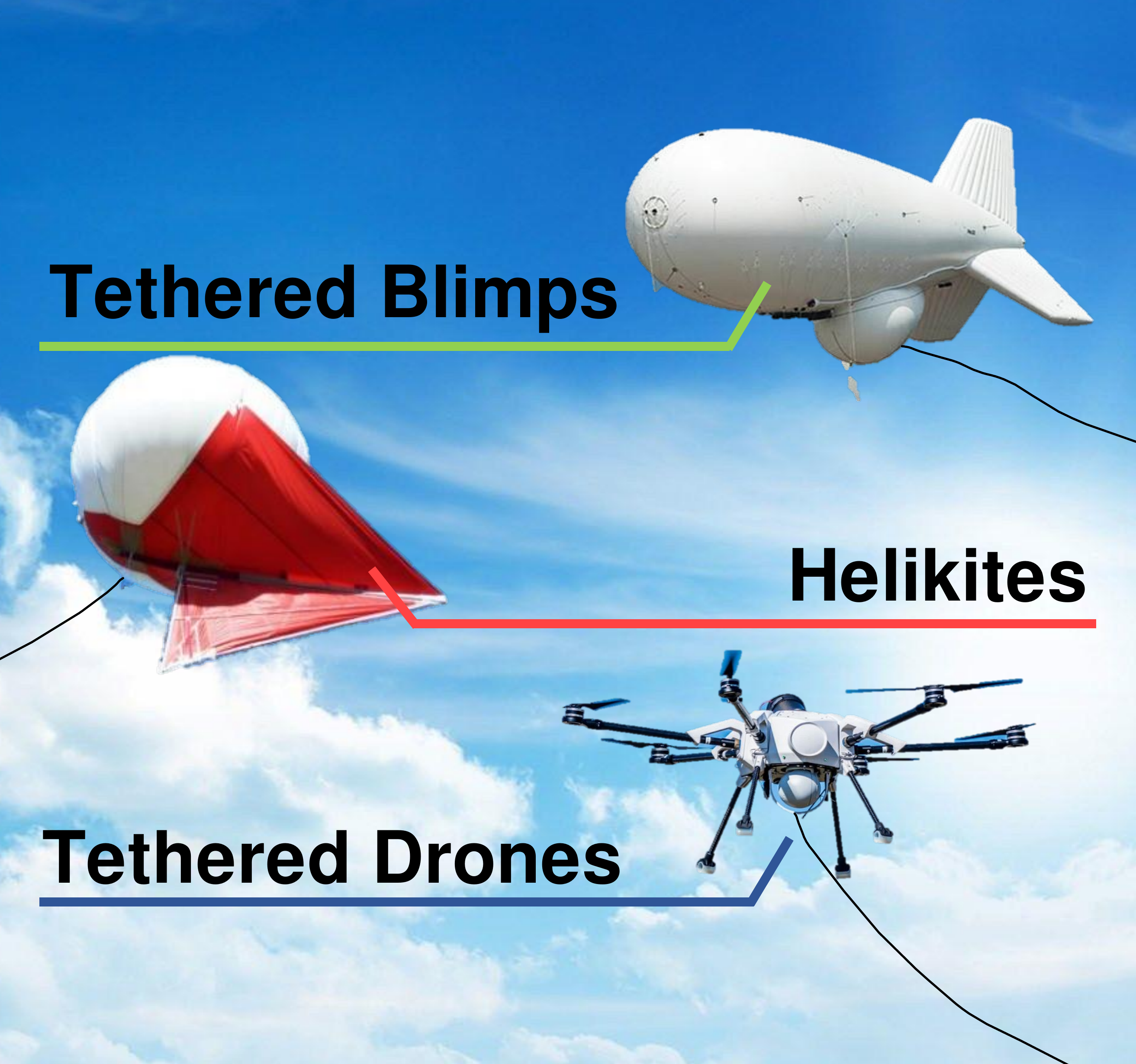}
\caption{The different types of NTFPs.}
\label{Fig.Type}
\end{figure}

\section{Overview of NTFP}

 \subsection{Types of NTFPs}
 
NTFPs are aerial vehicles that can fly in the air by opposing the force of gravity either by using a static or a dynamic lift while being tethered to the ground. For instance, blimps use static lift via lighter than air gas whereas drones use dynamic lift generated by motored engine. However, there are also hybrid NTFPs that use both static lift and dynamic lift such as Helikites. In the following, we describe only NTFPs that are relevant to our context shown in Fig.~\ref{Fig.Type}, since there are other type of NTFPs \cite{belmekki2020unleashing}.

\subsubsection{Tethered Blimps}
Blimps, also known as streamlined aerostat, are high performance platforms that sustain high speed wind, carry heavy payloads, and stay aloft at high altitudes over longer period of time. There are several categories of blimps, they differ in size, altitude, and payload. They can be used for tactical missions, operational missions, and strategic missions.

\subsubsection{Helikites}
Helikites are hybrid NTFPs that benefit from both static lift and dynamic lift. They are composed of an oblate-spheroid shaped balloon filled with helium to generate a static lift, and a kite structure to generate a dynamic lift. Consequently, they require less helium compared to the other comparably sized aerostats, and fly at  higher altitudes.

\subsubsection{Tethered Drones}
Tethered drones are drones with a physical link called tether that supply them with power and data. They usually have battery backup if the tether is damaged or if there is a cut in power. Since tethered drones have a constant supply of power, they can, in theory, stay up in the air for an unlimited period. In practice, they stay aloft for 2 to 4 days, after that the motor starts to heat up.

\begin{figure}[h!]
\centering
\includegraphics[scale=0.39]{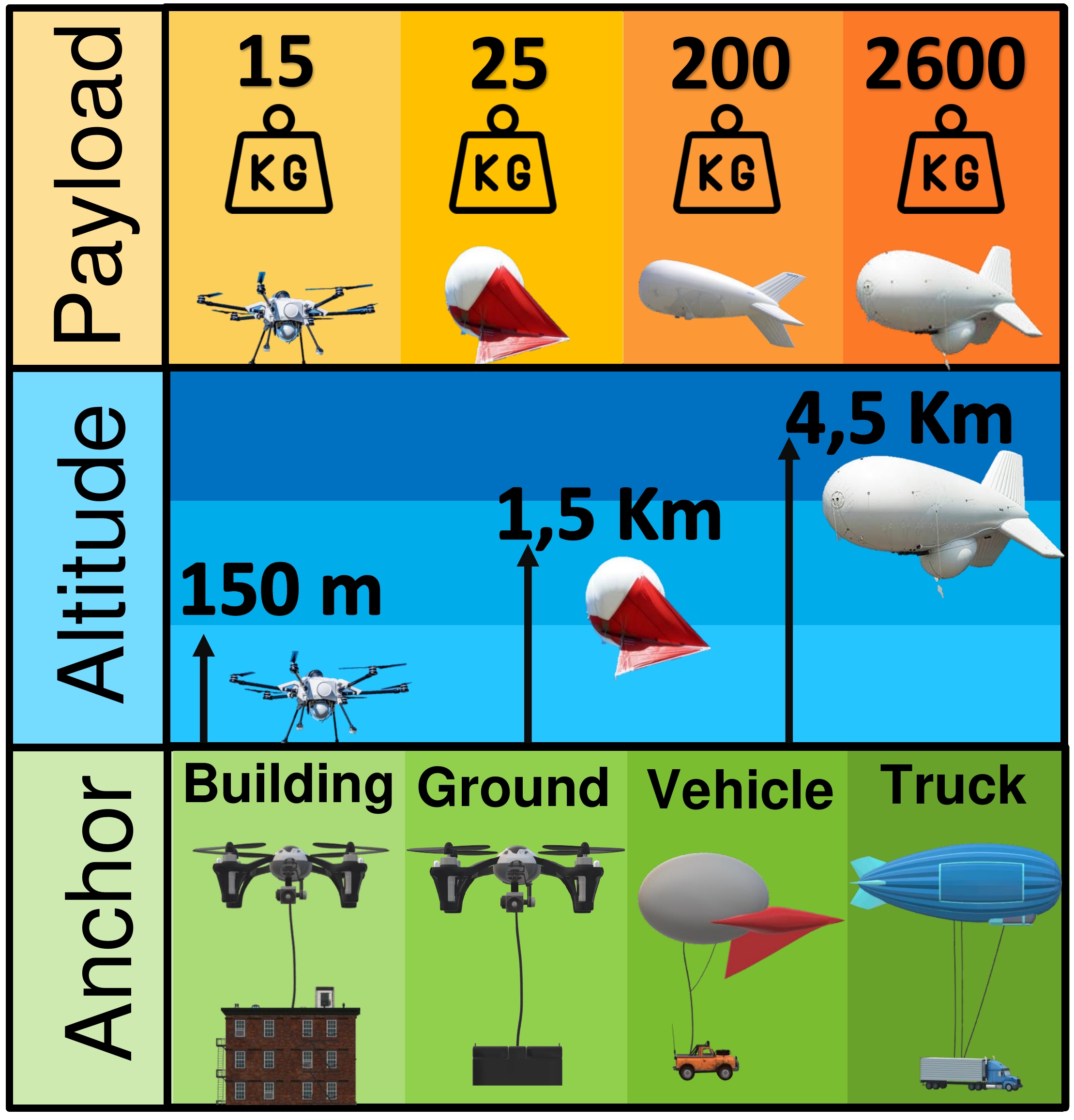}
\caption{Characteristics of NTFPs.}
\label{Fig.Charact}
\end{figure}

 \subsection{Components  of NTFPs}
\begin{itemize}
    \item Envelope: The envelope of NTFPs contains the gas that allows the platforms to soar and stay aloft
\item Lifting gas: The lifting gas, such as Helium, is a gas that has a lower density than the air, hence it permits to lift the envelope of the aerostat by creating buoyancy. 
Payload
\item The payload: The payload is the weight that the NTFP can carry while being in the air. Payload includes all the necessary equipment to operate the NTFP, power system, communication repeaters, etc.
\item Tether:
The tether maintains and stabilizes the NTFP to the ground when it floats in the air, it provides power to the NTFP through a power line, and it provides data to the NTFP through optical fibers.
\item Winches:
The winch is the device used to let out the tether during the launching process, adjust its tension
while the NTFP is aloft, and pull it in during the recovery process.
\item Anchor point:
The anchor point or anchor unit is the unit that the NTFP is anchored into. It maintains the NTFP in place while aloft.
\item Ground control station:
Ground control stations serve as operation base for NTFPs. They can be used to control the altitude of the NTFP, monitor the NTFP and the equipment they carry, store and process the data related to the mission,(e.g., videos and images).
\end{itemize}

 \subsection{Characteristics of NTFPs}
 The NTFPs have several characteristics related to their payload, altitudes and anchor point as shown in Fig.~\ref{Fig.Charact}. 
\begin{itemize}
    \item Payload:
NTFPs can carry different payloads depending on the type and the size of the NTFP. For instance, drones can carry payloads from 1~kg to 15~kg. Helikites can carry payload from 2~kg to 25~kg, and blimps carry payloads that ranges from 16~kg to 2600~kg.
    \item Altitudes:
NTFPs fly at different altitudes. Drones can fly at a maximum altitude of 150~m whereas Helikites can reach altitude of 1,5~km. Blimps, on the other hand, depending on their types, can fly at 250~m up to an altitude of 4,5~km.
    \item Anchor:
NTFPs can be anchored to different ground point which make them versatile and can be placed anywhere. They can be anchored on top of buildings, on grounds, on a vehicles, and on trucks. They can also be anchored on a ship or anchored on the sea.
\end{itemize}

\subsection{Main Advantages}

\begin{figure*}[ht!]
\centering
\includegraphics[scale=0.24]{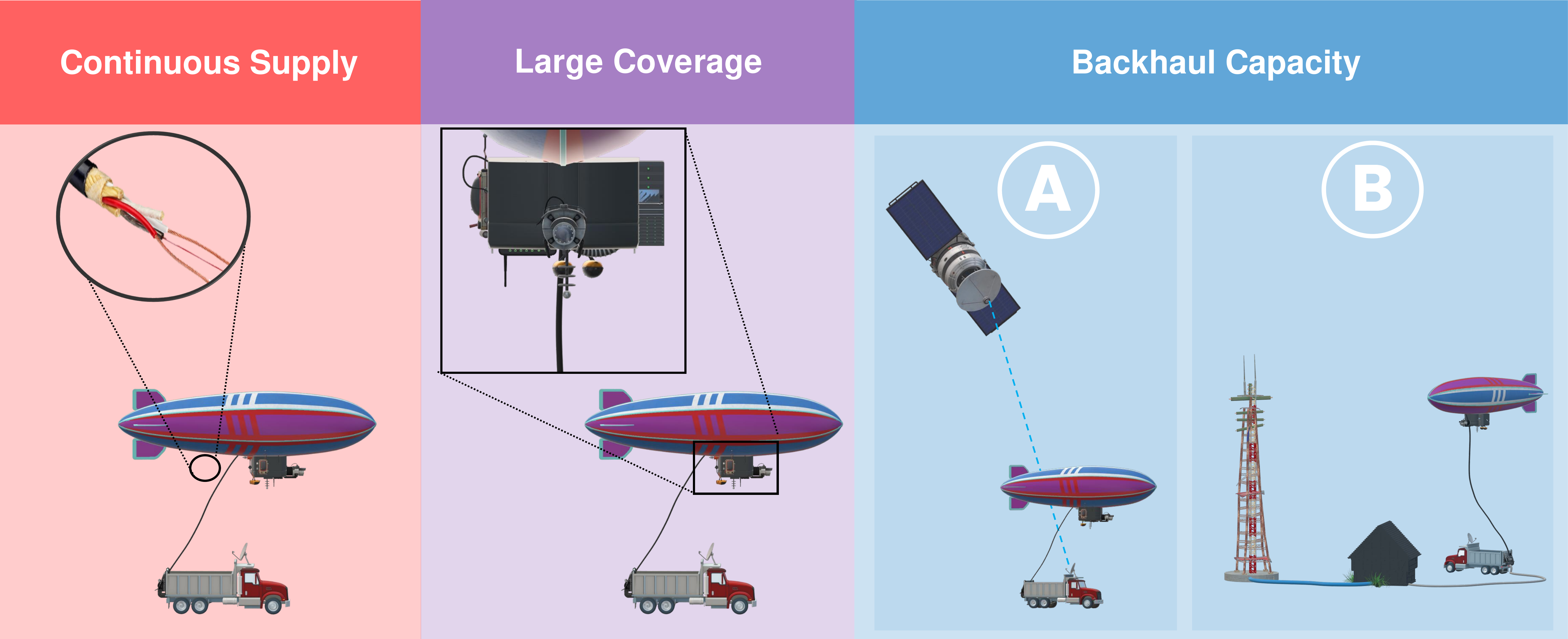} 
\caption{Main advantages of NTFPs.}
 \label{adv_NTFP}
\end{figure*}

\subsubsection{Continuous Supply}
One of the main advantages of NTFPs is that they benefit from a constant and continuous supply of data and power via their tether. The materiel that tethers are made from are usually synthetic fibers. Also, tethers are weather-proof, this allowing them to withstand high and low temperatures, humidity, rain, snow, lightning, and other harsh weather conditions.

\subsubsection{Large Coverage}
NTFPs, thanks to their high altitudes, have a larger communication coverage as well as visual coverage. Thus, allowing them to cover larger areas than cell towers, but also benefiting from a wide visual reach using cameras mounted as payload. This can also be extremely beneficial for surveillance mission. 

\subsubsection{Backhaul Capacity}
For fair comparison, NTFPs are compared with their free-flying counterparts such as drones and High-Altitude Platforms (HAPs). In that regard, NTFPs have a superior backhaul link capacity compared to free-flying platforms. In this case of free-flying platforms, their wireless backhaul is more prone to interference, hijacking, and suffers higher latency. In the contrary, tethered backhaul have a wired data-link via the tether allowing reliable, secure and high data rates communications. The backhaul link can be done via two methods as shown in Fig.~\ref{adv_NTFP} in Backhaul capacity, the method A and the method B. The method A uses a satellite backhaul between the NTFPs and a satellite. The method B uses a backhaul that is linked via a wire to an access point or communications infrastructure.

\begin{figure*}[th!]
\centering
\includegraphics[scale=0.53]{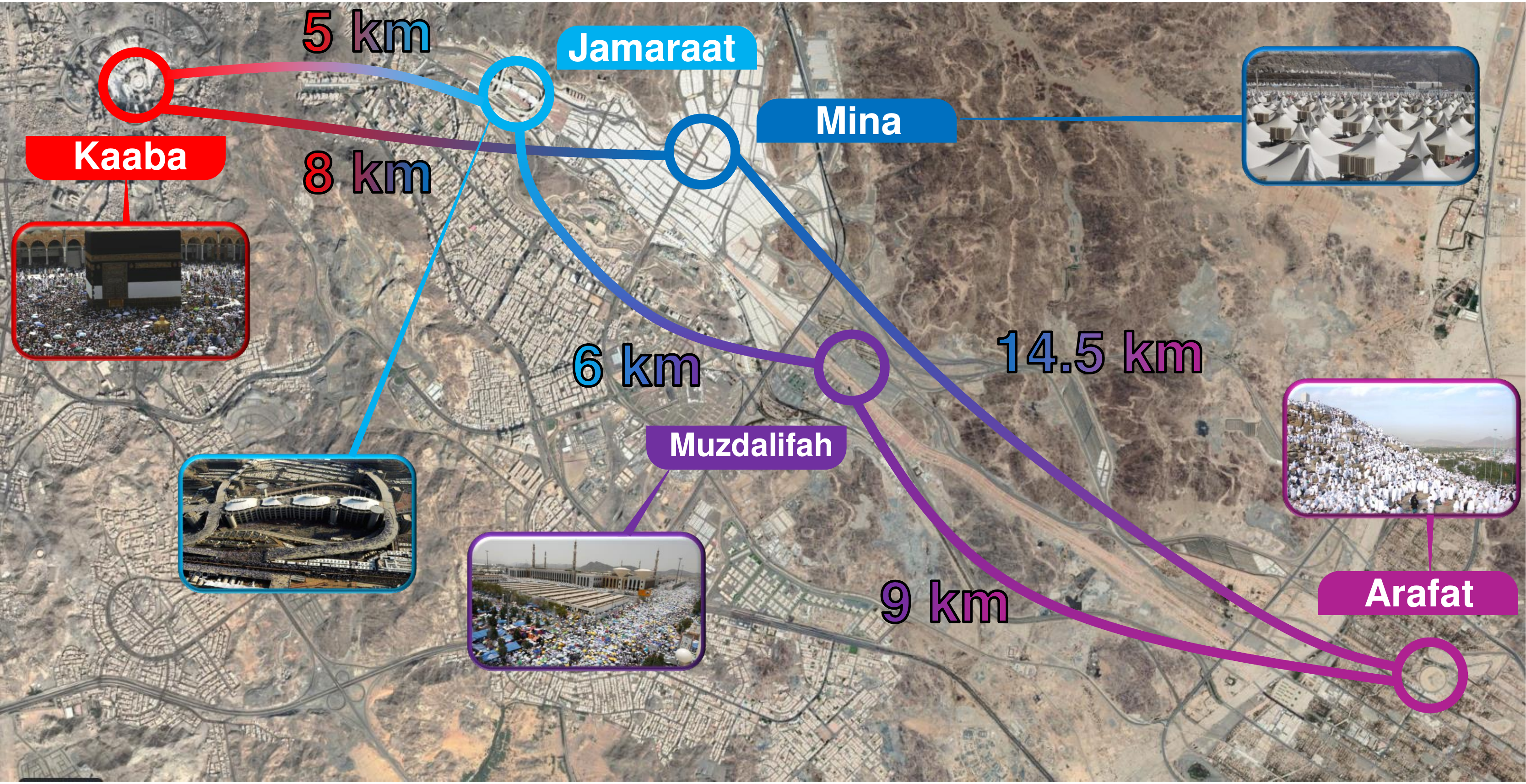}
\caption{The different locations in which the Hajj steps take place.}
\label{Fig.stepshajj}
\end{figure*}

\begin{figure}[h!]
\centering
\includegraphics[scale=0.3]{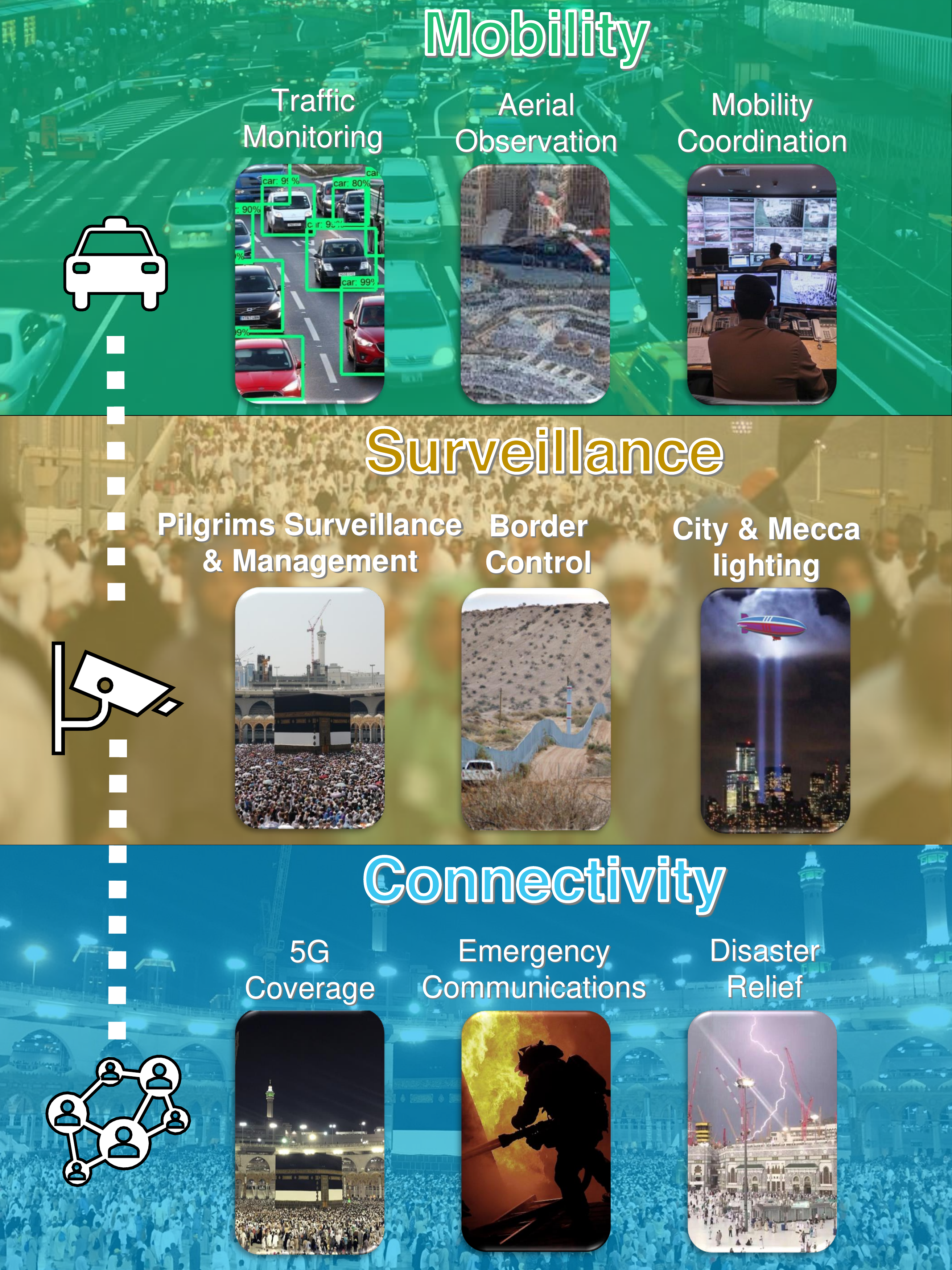}
\caption{Solutions offered by NTFPs during Hajj pilgrimage.}
\label{Fig.Appli}
\end{figure}

\begin{figure*}[h!]
\centering
\includegraphics[scale=0.6]{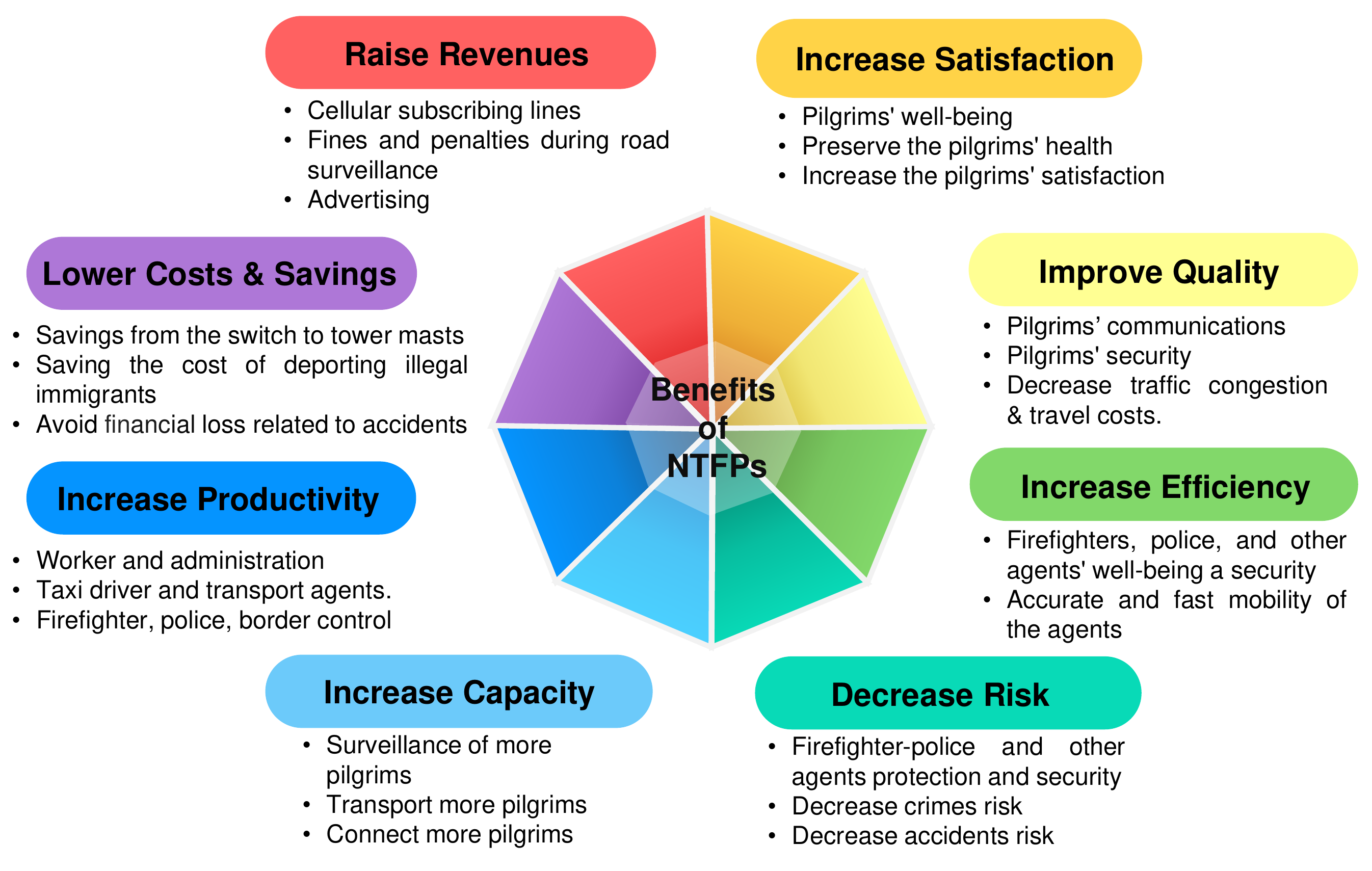}
\caption{Benefits of NTFPs in Hajj.}
\label{Fig.Bene}
\end{figure*}

\begin{figure}[h!]
\centering
\includegraphics[scale=0.4]{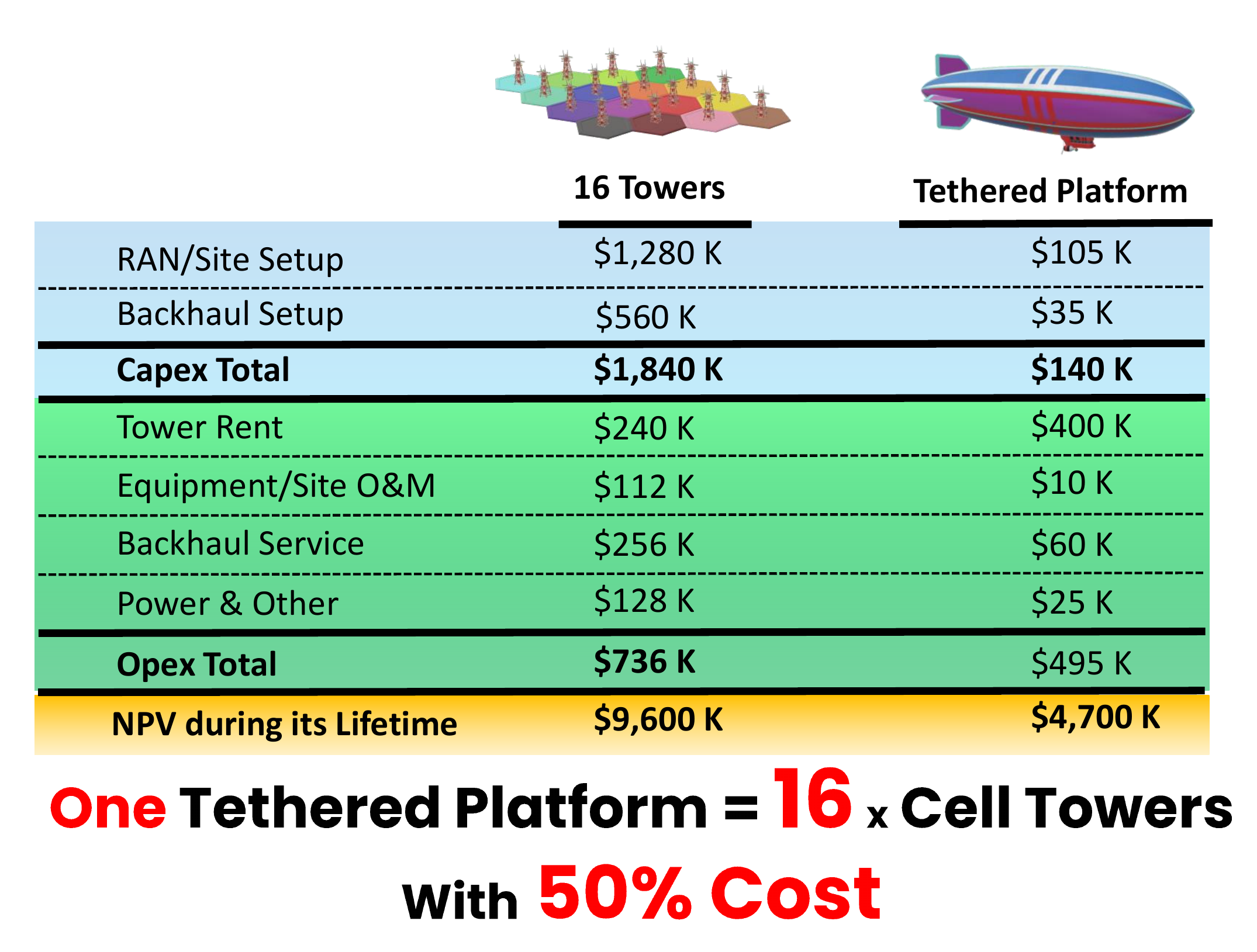}
\caption{Cost-analysis comparison between NTFPs and tower masts.}
\label{Fig.cost}
\end{figure}
\section{NTFP Solution for Hajj Pilgrimage}

\subsection{Key Locations During Hajj}
Before we present the applications of NTFPs in Hajj, we will briefly explain the major locations that pilgrims go to during their rituals \cite{isa}. The Hajj lasts five days, but pilgrims mostly arrive two-three days before hajj days start. The key locations are depicted in Fig.~\ref{Fig.stepshajj}.
On the first day, the pilgrims go from the Kaaba to Mina where they stay there until the next morning.
The second day, they head to Mount Arafat until sunset. After that, they go to Muzdalifah and spend the night there.
At the dawn of the third day, they leave for Mina again before the sunrise. The pilgrims go to Jamaraat.
Then they go back to circulate the Kaaba, and return to Mina to stay there for two-three days.
After that, pilgrims return to Makkah.

\subsection{Applications}
In this section, we will show the usage of NTFPs to solve the problems related to the Hajj and the resulting applications. The applications are related to mobility \& transportation, surveillance, and connectivity as shown in Fig.~\ref{Fig.Appli}.
\subsubsection{Mobility and Transportation}
First, most pilgrims come from outside Saudi Arabia, hence, they will arrive at the airport dedicated to the pilgrims. The road distance between the airport and Makkah is 95.5~km. Second, the pilgrims use transportation to reach the different locations in which the Hajj takes place. This highlights the importance of traffic fluidity, aerial observation, and mobility coordination of such a condensed large number of pilgrims in a relatively small area.

\begin{itemize}

  \item Traffic Monitoring: The cameras mounted on the NTFPs grant them a great view and coverage from the airport to Makkah and from Makkah to all the key locations. This will facilitate traffic monitoring by preventing possible road accidents, managing critical situations when accidents do occur, and detecting most of the illegal traffic violations.

  \item Aerial Observation: Although aerial observation falls into the category of surveillance, we use it here in the context of road transportation. Indeed, the aerial observation offered by NTFPs help track suspicious vehicles or follow target vehicles in runaway situations such as car chases. The large visual view provided by NTFPs allows the tracking of vehicles in a large area.

  \item Mobility Coordination: Having a great view of the city helps regulate the traffic lights according to the current situation in the road and also coordinating the different types of transportation for optimal traffic fluidity. It is also helpful to manage issues proactively by clearing or blocking some streets that are causing or may cause traffic congestion.
  
\end{itemize}
\subsubsection{Surveillance}
The most important aspect in Hajj is surveillance and the safety of pilgrims is paramount. Surveillance involves the pilgrims, illegal pilgrims, and criminals.
\begin{itemize}
\item Pilgrims Surveillance: A large number of pilgrims are concentrated in key locations (Kaaba, Mina, Arafat, Muzdalifah, and Jamaraat). NTFPs allow a constant surveillance over the pilgrims by preventing foreseeable accidents. Also, NTFPs allow a constant surveillance of any suspicious behavior or dangerous demeanor, and if that is the case, a notification is sent immediately to the security agents. 

\item Illegal Pilgrims: NTFPs have the ability to fly at high altitudes allowing them to watch the borders with a coverage of several kilometers. This will help the authorities to prevent pilgrims that do not possess Hajj permits from entering the areas restricted to the pilgrims. Additionally, the large visual coverage provided by NTFPs enable them to detect from a very long distance the illegal immigrants that try to enter the Saudi Arabia territory, and therefore, helping the border control agent to take adequate actions. 

\item City and Makkah lighting: NTFPs can be used to light the city from above, especially for the unlit areas. This will facilitate the surveillance and tracking of criminal activities such as car chases. Also, thanks to their night vision cameras, they can watch unlit areas, and light them if they detect suspicious behavior, and notify the ground agents.
\end{itemize}
\subsubsection{Connectivity}
Providing connectivity for a large number of pilgrims in a relatively small area during one week can be burdensome on the communication infrastructure, especially during peak hours.  This can be detrimental during the aftermath of a disaster.

\begin{itemize}

\item 4G/5G Coverage: NTFPs offer a wide and huge coverage thanks to their altitudes. The base station can be easily installed on the platforms without any changes making them technology-agnostic. Hence, providing the areas with seamless and high data rate internet connection to assist the existing communication infrastructures in the area. Also, NTFPs have a better line-of-sight clearance to assist not only Ground-based vehicular communications \cite{belmekkipaper}, but also aerial vehicular communications \cite{zaid2021technological}.

\item Temporary Coverage: After a disaster, NTFPs are the perfect solution to bring the coverage after the communications infrastructures are destroyed. They are easily and rapidly deployed. They can also help firefighters communicate between them for coordination during missions.

\item Advertising: In addition to the previous applications, NTFPs can also be used for advertising. Since they are up there in the sky, they are visible by everyone, and this can be a good opportunity for advertisers, but also to make important announcements during the Hajj.

\end{itemize}
\subsection{Benefits and Costs}
The benefits of usage of NTFP in Hajj are highlighted in Fig.~\ref{Fig.Bene}. In addtion, to highlight the cost-effectiveness of NTFPs, we will use as an example
the cost-analysis carried by Altaeros for their SuperTower blimp that provide coverage \cite{6GSummit}.
The SuperTower made by Altaeros is a high capacity and long-endurance NTFP, that delivers data and provides coverage \cite{Altaeroswebsite}. The analysis is shown in Fig.~\ref{Fig.cost}. First, we notice from Fig.~\ref{Fig.cost} that, in terms of coverage area, one NTFP flying at 250 m cover the same area as 16 tower masts. Second, we notice that
one NTFP has 90\% lower Capital Expenditure (CAPEX) and 30\% lower Operating Expenditure (OPEX) compared to 16 tower mast. Therefore, one NTFP flying at an altitude of 250 m cover the same area  as 16 tower masts at 50\% of the cost.


\begin{figure}[h!]
\centering
\includegraphics[scale=0.3]{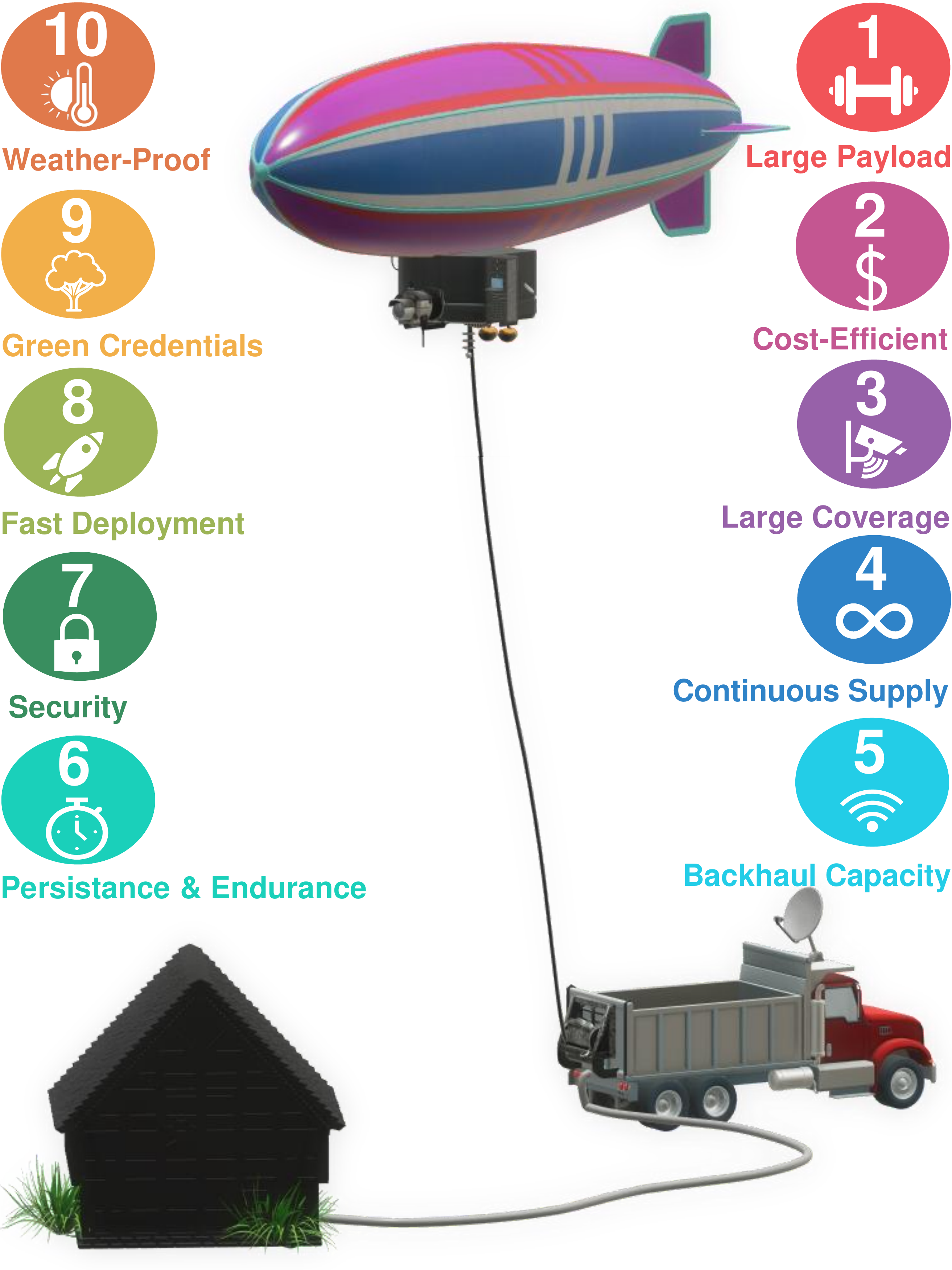}
\caption{Main advantages and benefits of NTFPs.}
\label{Fig.ADV}
\end{figure}
\begin{figure*}[h!]
\centering
\includegraphics[scale=0.3]{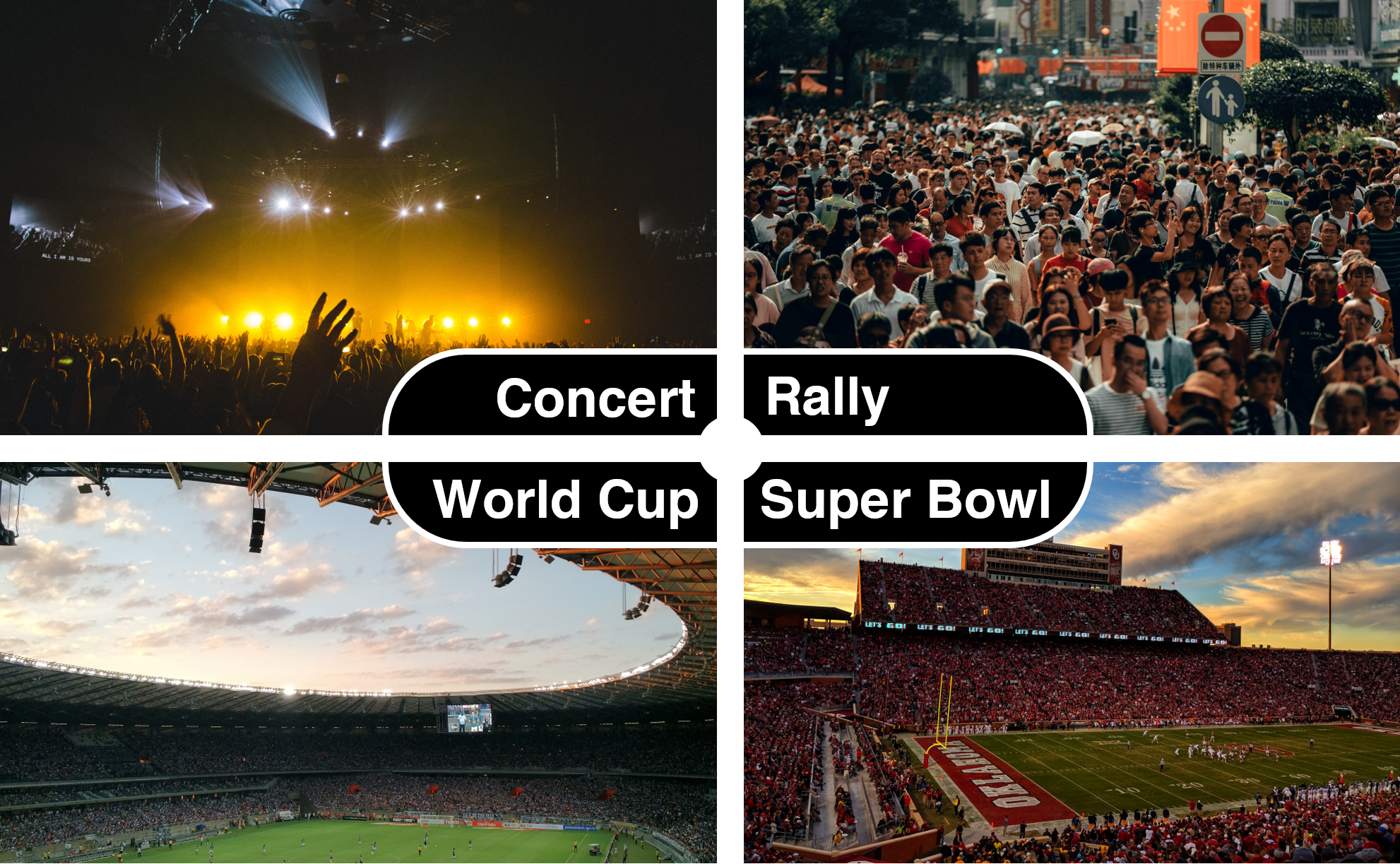}
\caption{The usage of NTFPs in major events.}
\label{Fig.}
\end{figure*}
\section{Additional Advantages and Case Studies of NTFPs}

\subsection{Additional Advantages}

In addition to the aforementioned advantages, here are other main benefits of NTFPs. Fig.~\ref{Fig.ADV} summarizes all the advantages of NTFPs. 

\subsubsection{Large Payload}
Another advantage of NTFPs is that they can carry high payloads that can reach 2600 kg. This allows them to carry all the necessary equipment to performs various tasks for different type of missions.

\subsubsection{Cost-Efficient}
NTFPs are cost-efficient compared to other communication platforms such as tower masts, satellite, and free-flying platforms. Additionally, they can perform multiple tasks by piggybacking on the existing platforms, such as surveillance, lighting, and recording.

\subsubsection{Persistence and Endurance}
NTFPs have high endurance, which is the time they can stay aloft, and high persistence, which the capability to stay steady in a given location for a prolonged period of time. These two features are necessary in surveillance and telecommunications missions.

\subsubsection{Fast Deployments}
NTFPs are quick to deploy, easy to reconfigure, and rapid to relocate if needed. Plus, their deployment, reconfiguration, and relocation can be handled by very few operators.

\subsubsection{Green Credentials}
NTFPs have a lower consumption of fuel and power compared the existing communication infrastructure. Especially tower masts and their exceptionally large fuel consumption.

\subsubsection{Weather-Proof}
One of the main strengths of NTFPs that their envelope and tether are weather-proof, granting them the ability to withstand extreme temperatures, humidity and other harsh weather conditions.

\subsection{Case Studies of NTFPs}
Although this paper discusses the usage of NTFPs in Hajj, this solution can also be used in different events that gather a massive number of people. In that regard, NTFPs can be very useful. We will present some of the other cases in which NTFPs can be integrated.

\subsubsection{Football World cup}
The football World Cup is an international football competition which is one of the most watch sport event in the world. This tournament takes place every four years, and last world cup took place in 2018 in Russia. According \cite{insidesport}, the World Cup host cities have received more than 5 million tourists. In this context, the usage of NTFPs can be beneficial in this type of sport events. Their usage can be used for surveillance such as hooligan riots or terrorist attacks. It can also enhance the coverage and the connectivity, especially in the location containing a large number of fans. Additionally, the high altitude of NTFPs can be used for broadcasting purposes.

\subsubsection{Olympic Games}
Olympic games are international sporting events that gathers thousands of athletes participating in a different sports. The Olympic games are held every four years in which more than 200 nations participate in them. In the 2016 Olympic games that were held in Rio, the mayor of Rio de Janeiro has stated that the city received 1.17 million tourists during the sporting events \cite{marca}. During these sporting events, integrating NTFPs can provide a large surveillance in different locations and connectivity. They can also be used to record the events from different altitude providing recording angles that are inaccessible with standard cameras.

\subsubsection{Super Bowl}
The super bowl is the National Football League (NFL) annual championship game. The number of fans that attend this event can reach more than 100,000 fans \cite{franchisesports}. The NTFPs can be used in the super bowl to enhance the surveillance in the stadium and in its vicinity. Also, they can be used to record the game and the musical Halftime Show from above. Finally, having a large number of people in a given location can slow the internet connection, hence, NTFPs will provide extra coverage during the super bowl event.

\subsubsection{Political Rallies}
Political rallies are gatherings that reunite people of similar political beliefs listening to a speaker. They are often scheduled during presidential elections. In these rallies, stampede can happen, and in some instances, causing deaths of the attendees \cite{France24}. 
This makes security an utmost concern during these events. NTFPs can be very helpful for surveillance purposes. In fact, having a large coverage of the area of interest and their surrounding allow a better assessment and anticipation of any suspect behavior. Hence, NTFPs can act as a deterrent. Additionally, they can be used to light the areas, since rallies can be held during nighttime.

\subsubsection{Concerts and Music Festivals}
Concerts and music festivals gather a large number of people. Usually, concerts seats anywhere from 5,000-10,000 people. Having such a large number of people in a same location and in a state of excitement (sometime under alcohol) can lead to accidents such as riot or stampede provoked by state of group panic. Also, excessive use of alcohol and lighters (which are common in concerts) can cause fire accidents and consequently suffocation accidents \cite{cheatsheet}. The usage of NTFPs can be extremely beneficial in these events. As stated in the above events, their usage can prevent terrorist attacks, help firefighters in the case of a fire accidents, or during defusing riots.

\section{Conclusion}
In this paper, we proposed an aerial-based solution using NTFPs to solve the problems related to Hajj. The NTFP solution solve problems related to mobility, such as traffic monitoring, aerial observation, and mobility coordination. But also problems related to security such as pilgrims surveillance, illegal pilgrims, and city lighting. Last, solution to problems related to connectivity such as cellular coverage, temporary coverage, and advertising are proposed. The benefits and advantages of the proposed solution are presented as well as the cost-efficiency of using NTFPs compared to the existing communication infrastructure, that is, tower masts. Finally, we showed how this solution can be used in other case studies involving large gatherings.

\bibliographystyle{ieeetr}
\bibliography{bibnoma}

\begin{thebibliography}{10}

\bibitem{britannica}
{Britannica}, ``{Hajj}.'' https://www.britannica.com/topic/{Hajj}.

\bibitem{saudigazette}
{Saudi Gazette}, ``Riyadh residents hope metro to end suffocating traffic
  jams.'' https://saudigazette.com.sa/article/501779.

\bibitem{khan2018analytical}
E.~A. Khan and M.~K.~Y. Shambour, ``An analytical study of mobile applications
  for {Hajj} and {Umrah} services,'' {\em Applied computing and informatics},
  vol.~14, no.~1, pp.~37--47, 2018.

\bibitem{majrashi2018user}
K.~Majrashi, ``User need and experience of {Hajj} mobile and ubiquitous
  systems: Designing for the largest religious annual gathering,'' {\em Cogent
  Engineering}, vol.~5, no.~1, p.~1480303, 2018.

\bibitem{alyami2020disaster}
A.~Alyami, C.~L. Dulong, M.~Z. Younis, and S.~Mansoor, ``Disaster preparedness
  in the {K}ingdom of {Saudi} {Arabia}: Exploring and evaluating the policy,
  legislative organisational arrangements particularly during the {Hajj}
  period,'' {\em European Journal of Environment and Public Health}, vol.~5,
  no.~1, p.~em0053, 2020.

\bibitem{thetealmango}
{The~Teal~Mango}, ``10 drones with longest flight time.''
  https://www.thetealmango.com/technology/drones-with-longest-flight-time/.

\bibitem{belmekki2020unleashing}
B.~E.~Y. Belmekki and M.-S. Alouini, ``Unleashing the potential of tethered
  networked flying platforms: Prospects, challenges, and applications,'' {\em
  arXiv preprint arXiv:2010.13509}, 2020.

\bibitem{isa}
{Islamic relief}, ``How to perform {Hajj} your complete step-by-step {Hajj}
  guide.''
  https://www.islamic-relief.org.uk/islamic-resources/hajj-in-islam/hajj-guide/.

\bibitem{belmekkipaper}
B.~E.~Y. Belmekki, A.~Hamza, and B.~Escrig, ``Cooperative vehicular
  communications at intersections over {Nakagami-$m$} fading channels,'' {\em
  Vehicular Communications}, vol.~19, p.~100165, 2019.

\bibitem{zaid2021technological}
A.~Abu~Zaid, B.~E.~Y. Belmekki, and M.-S. Alouini, ``Technological trends and
  key communication enablers for {eVTOLs},'' {\em arXiv preprint
  arXiv:2110.08830}, 2021.

\bibitem{6GSummit}
B.~Glass, ``Supertower: An introduction to tethered aerial cell towers.'' {6G}
  {Summit}, {Connecting the Unconnected}, 2020.

\bibitem{Altaeroswebsite}
Altaeros, ``Altaeros website.'' {http://www.altaeros.com/}.

\bibitem{insidesport}
{Insidesport}, ``{FIFA} {World Cup} 2018 generates 5 million tourists for
  {Russia}.''
  https://www.insidesport.co/fifa-world-cup-2018-generates-5-million-tourists-for-russia/.

\bibitem{marca}
{Marca}, ``Rio de janeiro welcomed 1.17 million tourists in two weeks.''
  https://www.marca.com/en/olympic-games/2016/08/24/57bda7a0468aeb3e158b4596.html.

\bibitem{franchisesports}
{Franchise Sports}, ``Top 10 super bowl attendances of all-time.''
  https://franchisesports.co.uk/biggest-super-bowl-attendances/.

\bibitem{France24}
{France24}, ``Crowd stampedes at presidential rally.''
  https://www.france24.com/en/20110213-rally-president-goodluck-jonathan-turns-deadly-crowd-stampede-harcourt-nigeria.

\bibitem{cheatsheet}
{Cheatsheet}, ``10 concerts that ended in tragedy.''
  https://www.cheatsheet.com/entertainment/6-concerts-that-ended-in-tragedy.html/.

\end{thebibliography}

\begin{IEEEbiography}
    [{\includegraphics[width=2.6cm,height=2.8cm,clip,keepaspectratio]{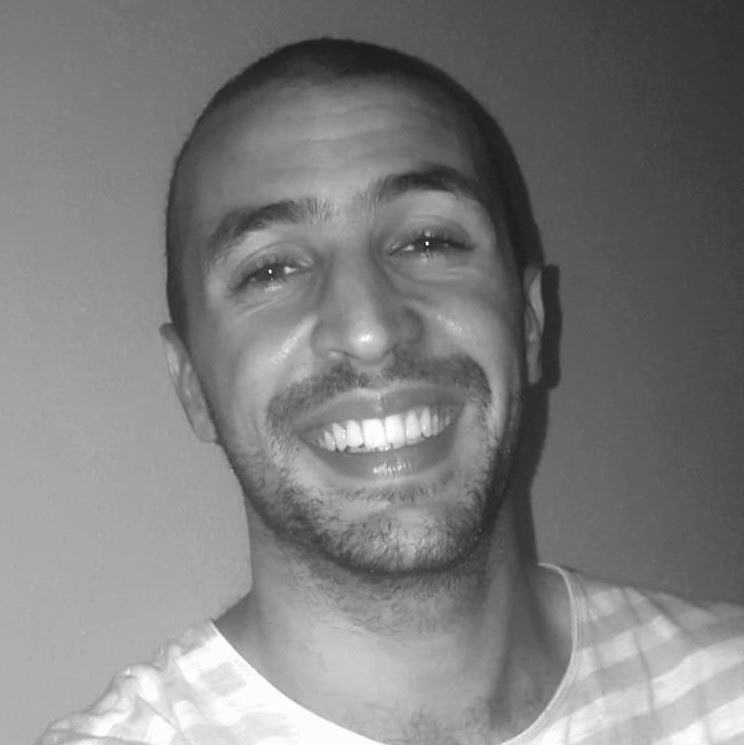}}]{Baha Eddine Youcef Belmekki}
received the B.S. degree in Electronics, and the M.Sc. degree in Wireless Communications and Networking from University of Science and Technology Houari Boumediene, Algiers, Algeria, in 2011 and 2013, respectively. From 2013 to 2014, he was with the Department of Radio Access Network, Huawei Technologies, Algiers, Algeria. From 2014 to 2016, he was an Assistant Professor with University of Science and Technology Houari Boumediene, Algiers, Algeria. He obtained his Ph.D. degrees in Wireless Communications and Signal Processing from the National Polytechnic Institute of Toulouse (INPT) in 2020. 
From 2019 to 2021, he worked as teaching and research assistant at the National Polytechnic Institute of Toulouse, France.
He is currently a postdoctoral research fellow at the Communication Theory Laboratory of King Abdullah University of Science and  Technology (KAUST), Thuwal, Makkah Province, Saudi Arabia.
His research interests include non-orthogonal multiple access systems and stochastic geometry analysis of vehicular and aerial networks.
\end{IEEEbiography}
\vspace{-1cm}

\begin{IEEEbiography}
    [{\includegraphics[width=2.6cm,height=2.8cm,clip,keepaspectratio]{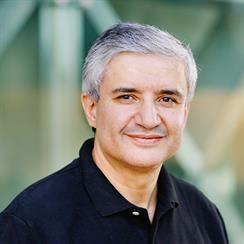}}]
    {Mohamed-Slim Alouini}
(S’94-M’98-SM’03-
F’09) was born in Tunis, Tunisia. He received
the Ph.D. degree in Electrical Engineering from
the California Institute of Technology (Caltech),
Pasadena, CA, USA, in 1998. He served as a
faculty member in the University of Minnesota,
Minneapolis, MN, USA, then in the Texas A\&M
University at Qatar, Education City, Doha, Qatar
before joining King Abdullah University of
Science and Technology (KAUST), Thuwal,
Makkah Province, Saudi Arabia as a Professor
of Electrical and Computer Engineering in 2009. His current research
interests include the modeling, design, and performance analysis of wireless
communication systems.
\end{IEEEbiography}
%



\end{document}